\documentclass[useAMS,usenatbib]{mnras}
\usepackage{graphicx} 
\usepackage{subcaption}
\usepackage{amsmath} 
\usepackage{xcolor}
\usepackage{float}
\usepackage{caption}
\usepackage{placeins}
\usepackage{lipsum}
\usepackage{multicol}
\usepackage{soul}
\usepackage{amssymb}
\usepackage{ragged2e}
\usepackage{threeparttable}
\usepackage{comment}
\def\refjnl#1{{\rm#1}}
\def\araa{\refjnl{ARA\&A}}             
\def\apj{\refjnl{ApJ}}                 
\def\apjl{\refjnl{ApJ}}                
\def\apjs{\refjnl{ApJS}}               
\def\ao{\refjnl{Appl.~Opt.}}           
\def\aap{\refjnl{A\&A}}                
\def\mnras{\refjnl{MNRAS}}             
\def\prd{\refjnl{Phys.~Rev.~D}}        
\def\pasa{\refjnl{PASA}}               
\def\nat{\refjnl{Nature}}              
\def\physrep{\refjnl{Phys.~Rep.}}   

\def\be{\begin{equation}} 
\def\ee{\end{equation}}

\def\gsim{\lower.5ex\hbox{\gtsima}} 
\def\lsim{\lower.5ex\hbox{\ltsima}} \def\gtsima{$\; \buildrel > \over 
\sim \;$} \def\ltsima{$\; \buildrel < \over \sim \;$} \def\prosima{$\; 
\buildrel \propto \over \sim \;$} \def\gsim{\lower.5ex\hbox{\gtsima}} 
\def\lsim{\lower.5ex\hbox{\ltsima}} 
\def\simgt{\lower.5ex\hbox{\gtsima}} 
\def\simlt{\lower.5ex\hbox{\ltsima}} 
\def\simpr{\lower.5ex\hbox{\prosima}} \def\la{\lsim} \def\ga{\gsim} 
  
 \def\gtsima{$\; \buildrel > \over \sim \;$} 
\def\ltsima{$\; \buildrel < \over \sim \;$} 
\def\gsim{\lower.5ex\hbox{\gtsima}} 
\def\lsim{\lower.5ex\hbox{\ltsima}} 
\def\simgt{\lower.5ex\hbox{\gtsima}} 
\def\simlt{\lower.5ex\hbox{\ltsima}} 
\def\simpr{\lower.5ex\hbox{\prosima}} 
\def\la{\lsim} 
\def\ga{\gsim}

\def\msun{\,{\rm \Msun}}

\def\E3{{\cal E}_{\rm g}^{III}}

\def\Msun{\rm M_\odot}

\def\Msun{\rm M_\odot}

\def\M*{M_*}

\def\Z*{Z_*}
\def\L*{L_*}
\def\muv{\rm M_{UV}}

\def\fs{f_*}
\def\fej{f_*^{\rm{ej}}}
\def\feff{f_*^{\rm{eff}}}

\def\der{{\rm d}}

\newcommand\code[1]{\textsc{\MakeLowercase{#1}}}

\def \fescuv{f_{\rm{esc}}^{\rm{UV}}}

\title[21cm global signal in JWST and ALMA era]{Predictions of the 21cm global signal in the JWST and ALMA era}

\author[Chatterjee et al.]{
Atrideb Chatterjee,$^{1}$\thanks{E-mail:atrideb.chatterjee@iucaa.in }
Pratika Dayal$^{2}$ \&
Valentin Mauerhofer$^{2}$ 
\\
$^{1}$Inter-University Centre for Astronomy and Astrophysics, Post Bag 4, Ganeshkhind, Pune 411007, India\\
$^{2}$ Kapteyn Astronomical Institute, University of Groningen, P.O. Box 800, 9700 AV Groningen, The Netherlands
}

\date{Accepted XXX. Received YYY; in original form ZZZ}

\pubyear{2023}

\begin{document}
\label{firstpage}
\pagerange{\pageref{firstpage}--\pageref{lastpage}}
\maketitle

\begin{abstract}
We calculate the redshift evolution of the global 21cm signal in the first billion years using an advanced semi-analytic galaxy formation model \code{delphi}.
Employing only two redshift- and mass-independent free parameters, our model predicts galaxy populations in accord with data from both the James Webb Space Telescope (JWST) and the Atacama Large Millimetre Array (ALMA) at $z \sim 5-12$. In addition to this ``fiducial" model, which fully incorporates the impact of dust attenuation, we also explore an unphysical ``maximal" model wherein galaxies can convert a 100\% of their gas into stars instantaneously (and supernova feedback is ignored) required to explain JWST data at $z >=13$. We also explore a wide range of values for our {\it 21cm} parameters that include the impact of X-ray heating ($f_{\rm X,h} =0.02-2.0$) and the escape fraction of Lyman Alpha photons ($f_\alpha = 0.01-1.0$). Our key findings are: {\it (i)} the fiducial model predicts a global 21cm signal which reaches a minimum brightness temperature of $ T_{\rm b, min}\sim -215$ mK at a redshift $z_{\rm min} \sim 14$; {\it (ii)} since the impact of dust on galaxy properties 
only becomes relevant at $z <= 8$, dust does not have a sensible impact on the global 21cm signal; {\it (iii)} the ``maximal" model predicts $T_{\rm b, min}= -210$ mK as early as $z_{\rm min} \sim 18$; {\it (iv)} {\it galaxy formation} and {\it 21cm parameters} have a degenerate impact on the global 21cm signal. A combination of the minimum temperature and its redshift will therefore be crucial in constraining galaxy formation parameters and their coupling to the 21cm signal at these early epochs.

\end{abstract}




\section{Introduction}
The sky-averaged 21cm global signal, arising from the hyperfine transition of neutral hydrogen, is an excellent probe of the era of cosmic dawn \citep[e.g.][]{pritchard2012, 10.1088/2514-3433/ab4a73}. The redshift evolution and the depth of the absorption trough of this signal can be used to infer the properties of the first generation of stars \citep{2020MNRAS.496.1445C, 2020MNRAS.497.2839L, 2022MNRAS.516..841G, 2023MNRAS.520.3609V}, obtain constraints on the highly uncertain astrophysical parameters used in semi-numerical models of the high-redshift Universe \citep{2015MNRAS.449.4246G, 2017ApJ...848...23K, 2018MNRAS.477.3217G, 2019MNRAS.484..282G, 2021MNRAS.503.4551G, 2021MNRAS.507.2405C} and even obtain hints on the elusive nature of dark matter \citep{2019PhRvD.100l3005B, 2019MNRAS.487.3560C, 2022PhRvD.105h3011G, 2022PhRvD.106f3504F}, to name a few. In 2018, the EDGES (Experiment to Detect the Global Epoch of Reionization Signature) collaboration \citep{bowman2018} made the first detection \citep[see, e.g.][]{2018Natur.564E..32H, 2019ApJ...874..153B, 2019ApJ...880...26S, 2020MNRAS.492...22S} of this signal which has recently been ruled out at a $95.3\%$ confidence limit by the Shaped Antenna Measurement of the Background Radio Spectrum 3  \citep[SARAS-3;][]{2022NatAs...6..607S}. 

Understanding the emergence of the first sources, that can heat up and ionize neutral hydrogen \citep[for a review see e.g.][]{dayal2018}, are therefore crucial in making realistic predictions regarding the shape and amplitude of the 21cm signal during the first billion years. We are currently in a golden era for the hunt for such sources, driven by a combination of space- and ground-based facilities, including the Hubble Space Telescope (HST), the UK Infra Red Telescope (UKIRT), the Very Large Telescope (VLT), the Subaru telescope, 
and most recently, the James Webb Space Telescope (JWST) and the Atacama Large Millimetre Array (ALMA). These have been used to calculate galaxy properties, including their ultra-violet (UV) luminosities, stellar and dust masses and even morphologies well within the first billion years. The JWST, in particular, is allowing unprecedented observations of early galaxy formation, yielding a number of galaxy candidates between $z \sim 9-16.5$ \citep{bradley2022, donnan2023, atek2022, naidu2022a, adams2022} although caution must be exerted at $z \gsim 12$ where the redshift and nature of the sources remain debated pending spectroscopic confirmations \citep{adams2022, naidu2022, haro2023}. These observations are supplemented by ALMA programs yielding dust masses for galaxies at redshifts as high as $z \sim 5-7.5$ \citep[e.g.][]{bethermin2020, bouwens2022, inami2022}. ALMA observations seem to indicate that as much as $30-60\%$ of the star formation rate could be missed in the UV at $z \sim 7$ due to dust attenuation in bright sources \citep{Algera2023a}, rendering it crucial to account for the impact of dust even at these early epochs. 

Given its implications, a number of works have aimed at obtaining the 21cm global signal based on different underlying galaxy formation models \citep{2010ascl.soft10025S, 2011MNRAS.411..955M, 2014MNRAS.437L..36F, 2017MNRAS.472.4508S, 2018MNRAS.477.1549H,  2018MNRAS.476.1741G, 2020MNRAS.498.6083E, 2023arXiv230511441Y} with some beginning to use the latest JWST constraints \citep[e.g.][]{hassan2023}. In this work, we improve on such previous calculations by using galaxy populations that are fully base-lined against all available galaxy data-sets, including their dust attenuations and masses, at $z \sim 5-9$ using only two {\it redshift- and mass-independent parameters} \citep{dayal2022, mauerhofer2023}. The key aim of this work is to predict a 21cm global signal from this realistic galaxy population, covering all of the physically-plausible free-parameter space. This endeavour is crucial for instruments that aim to observe the 21cm global signal including SARAS-3 \citep{2022NatAs...6..607S}, the Large-Aperture Experiment to Detect the Dark Ages \citep[LEDA;][]{Greenhill_2012}, SCI-HI \citep{Voytek_2014}, the Broadband Instrument for Global hydrogen Reionisation Signal \citep[BIGHORNS,][]{Sokolowski_2015}, the Radio Experiment for the Analysis of Cosmic Hydrogen \citep[REACH,][]{2022JAI....1150001C} and the Cosmic Twilight Polarimeter \citep[CTP;][]{Nahn_2018}.

Throughout this paper, we adopt a $\Lambda$CDM model with dark energy, dark matter and baryonic densities in units of the critical density as $\Omega_{\Lambda}= 0.691$, $\Omega_{m}= 0.308$ and $\Omega_{b}= 0.049$, respectively, a Hubble constant $H_0=100\, h\,{\rm km}\,{\rm s}^{-1}\,{\rm Mpc}^{-1}$ with $h=0.67$, spectral index $n=0.96$ and normalisation $\sigma_{8}=0.81$ \citep[][]{planck2016}.
\section{The theoretical Model}
The theoretical model used in this work combines results from a state-of-the-art semi-analytic galaxy formation model with an analytic formalism for calculating the global 21cm signal, as detailed in what follows. 


\subsection{The semi-analytic galaxy formation model}
\label{sam}
In order to model the high-redshift galaxy population, we use the \code{delphi} semi-analytical model which jointly tracks the assembly of dark matter halos and their baryonic components at $z \gsim 5$; interested readers are referred to our previous works \citep{dayal2014a, dayal2022, mauerhofer2023} for complete details. We start by building dark matter halo merger trees for 600 halos at redshift $z=4.5$ uniformly distributed (in log space) between $\log(M_h/\Msun) = 8-14$ using a binary merger tree algorithm  \citep{parkinson2008}. The assembly history of these halos is then tracked up to a maximum redshift of $z \sim 40$ with a time resolution of 30 Myr and a mass resolution of $10^8 \msun$. 

In terms of baryonic physics, the gas mass ($M_g$) in starting halos (i.e. halos that have no progenitors) is assumed to be proportional to the dark matter mass following the cosmological ratio such that $M_g = (\Omega_b/\Omega_m) M_h$;  for halos that have progenitors, the total gas mass is the sum of that brought in by merging progenitors and that smoothly-accreted from the intergalactic medium (IGM). The available gas mass is assumed to form stars with an ``effective efficiency" of $\feff$, which is the minimum between the efficiency that produces enough Type II Supernova (SNII) energy to eject the remainder of the gas ($\fej$) and an upper maximum (mass- and redshift- independent) threshold ($\fs$) i.e. $\feff = min[\fej,\fs]$. The upper limit ($\fs$) is essentially driven by observations of the evolving ultra-violet luminosity function at $z \sim 5-12$ that seems to indicate a constant efficiency of star formation for galaxies more massive than about $10^{9.5}\msun$ in terms of halo mass \citep[see e.g.][]{dayal2014a,mauerhofer2023}. The value of $\fej$ naturally depends on the fraction of SNII energy that can couple to the gas ($f_w$); we use this formalism to derive the gas mass ejected at any time step due to SNII feedback. At any time step this results in a newly formed stellar mass of $M_* = \feff M_g$ and an associated continuous star formation rate of $\psi = M_*/30{\rm Myr}$. We compute the intrinsic UV emission, $L_{\rm{UV}}^{\rm{int}}$, using the \code{bpass}  (v2.2.1) stellar population synthesis model \citep{BPASS1,BPASS2} which assumes a Kroupa IMF \citep{Kroupa01} between $0.1-100\msun$. As an example, a galaxy with metallicity of $7.5\%~\rm{Z}_{\odot}$ and $\psi = 1\msun~{\rm yr^{-1}}$ has a value of $L_{\rm{UV}}^{\rm{int}} = 1.23\times 10^{40} [\rm{erg~ s^{-1}\AA^{-1}}]$ and an ionising photon production rate of $\rm{\dot{N}_{ion}} = 2.83 \times 10^{53} [\rm{s^{-1}}]$.

\begin{figure}
    \centering
    \includegraphics[width=0.5\textwidth]{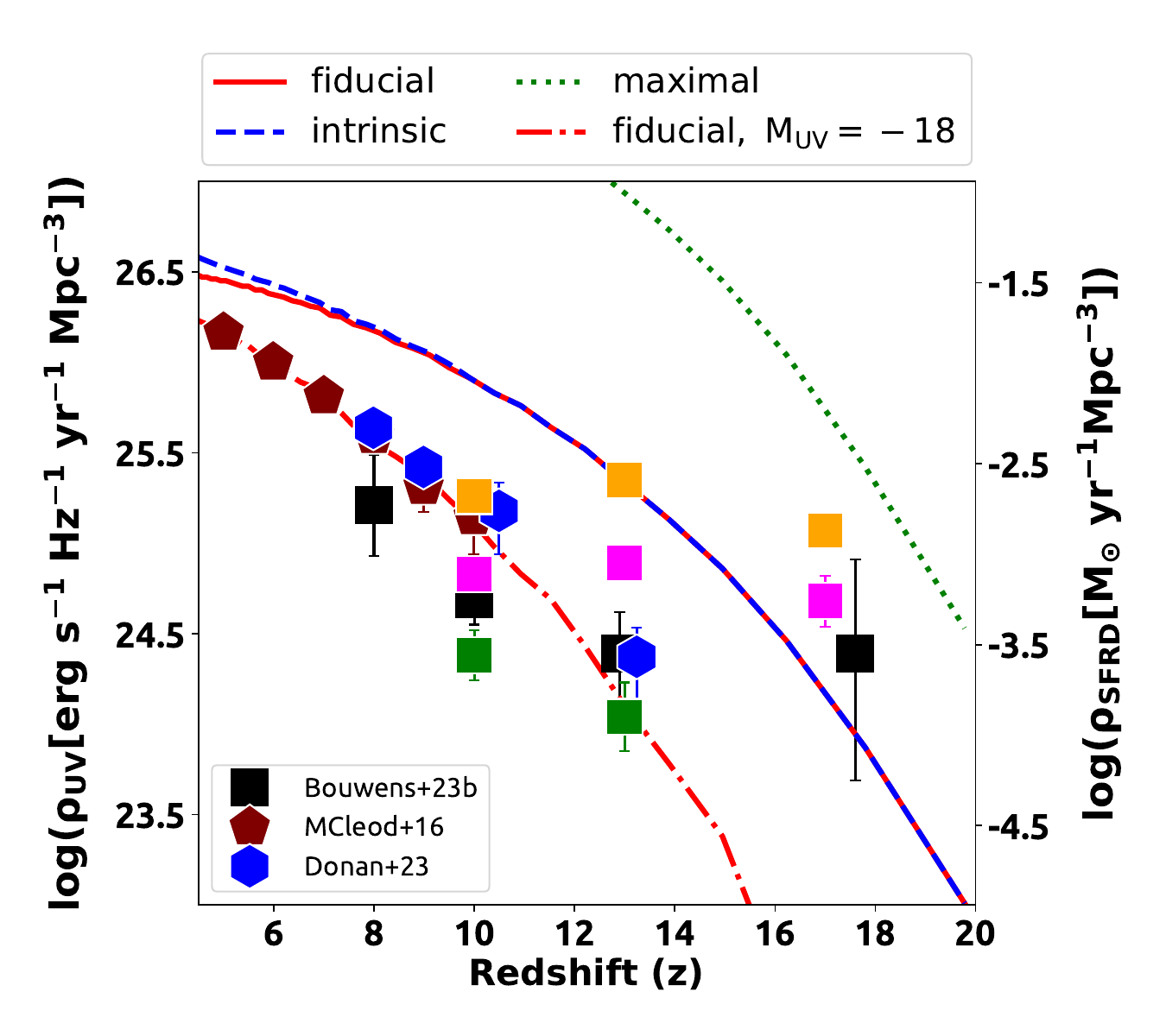}
    \caption{ The redshift evolution of the UV luminosity density and the corresponding star formation rate density at $z \sim 5-20$. As marked, the dashed (blue) and solid (red) lines show results for the intrinsic (no dust attenuation) and the fiducial (with dust attenuation) models considering all galaxies; the dot-dashed (red) line shows results for the fiducial model for galaxies brighter than $\muv=-18$. Finally, the dotted (green) line shows the results for the ``maximal" model. The points show observational results, as marked, from \citet[][diamonds]{donnan2023} who use a magnitude threshold of $\muv = -17$, \citet[][pentagons]{McLeod2016} who use a magnitude threshold of $\muv = -17.7$ and \citet[][squares; black for fiducial, and green, magenta and orange for ``robust", ``solid" and ``possible" literature detections, using a magnitude threshold of $\muv = -19$, respectively]{Bouwens23_solid}.  }
    \label{fig:uvld}
\end{figure}

 Assuming perfect mixing of gas, metals and dust at each time-step, we also include the key processes of dust and metal production, astration (into star formation), destruction (of dust into metals), ejection (of metal and dust) and dust grain growth in the interstellar medium (ISM; that leads to a corresponding decrease in the metal mass). While we use the latest mass- and metallicity-dependent stellar metal yields between $1-50~\Msun$ \citep{kobayashi2020}, we assume each SNII to form $0.5\msun$ of dust \citep{dayal2022}. We use a dust distribution radius ($r_d$) that is equal to the gas radius ($r_{\rm{gas}}$) that scales as $r_d= r_{\rm{gas}} =  0.18[(1+z)/7] r_{\rm{vir}}$, where $r_{\rm{vir}}$ is the virial radius. After assuming a grain size of $a=0.05 \mu m$ and a material density of $s = 2.25 \rm{g \, cm^{-3}}$, the dust optical depth is calculated as $\tau_d = 3 M_d [4 \pi r_d^2 a s]^{-1}$. The corresponding fraction of UV photons that can escape the galaxy (the UV escape fraction) is then calculated as $\fescuv = (1 - e^{-\tau_d})[\tau_d]^{-1}$. The UV luminosity that is ``observed", accounting for this dust attenuation, is then 
$L_{\rm{UV}}^{\rm{obs}} = \fescuv L_{\rm{UV}}^{\rm{int}}.$  

Matching to galaxy observations (including the evolving UV luminosity function, stellar mass function and their derived quantities) at $z \sim 5-9$ requires free parameter values of $f_*=0.15$ and $f_w=0.06$ - this is the {\it fiducial} galaxy formation model used throughout this work. As discussed in \citet{mauerhofer2023}, this model yields observables that are in good accord with the data up to $z \sim 12$. 

We start by showing a comparison of the UV luminosity density obtained from the model with observations in Fig. \ref{fig:uvld}. We use a conversion factor of $\kappa_{\rm UV} =  \psi/L_{\rm{UV}}= 1.15 \times 10^{-28}$ [$\rm{M}_{\odot} \rm{yr}^{-1} / \rm{erg} \, \rm{s}^{-1} \rm{Hz}^{-1}$] to obtain a corresponding star formation rate density from the UV luminosity density \citep{madau2014}, as shown in the same figure. Firstly, as a validation of the star formation rate density of the model, we show that considering galaxies brighter than $\muv =-18$, the fiducial model predicts a UV luminosity density that is in excellent agreement with the observations (which use luminosity limits ranging between $\muv=-17$ and $-19$) at $z \lsim 13$. Secondly, we find that considering all galaxies, the UV luminosity density predicted by both the intrinsic (no dust attenuation) and fiducial (including dust attenuation) models only show a visible difference at $z \lsim 8$ where the impact of dust attenuation becomes important. As might be expected, these models that include all galaxies predict significantly higher UV luminosity densities as compared to the observations - by a factor 2 at $z \sim 5$ increasing to an order of magnitude by $z \sim 13$. Finally, we find that the ``solid" and ``possible" data points inferred by \citet{Bouwens23_solid} at $z \gsim 13$ lie above the UV luminosity density predicted by our model, even considering all galaxies. Although spectroscopic confirmations will be crucial in validating the high-redshift nature of these sources, we also calculate the {\it ``maximal"} star formation rates (and the associated UV luminosities) allowed by our model at $z \gsim 12$, assuming no feedback (i.e. $f_w=0$) and a star formation efficiency of $\feff=1.0$. Encompassing the observations, this ``maximal" model provides an upper limit to the theoretical UV luminosity density. We now use the fiducial, intrinsic and maximal models (summarised in Table \ref{table1}), considering all galaxies, to calculate the global 21cm signal in what follows.

\subsection{Calculating the 21cm Global signal}
\label{sec:evolution_21cm}
\begin{table}
\caption {For the model in column 1, we note the free parameter values including the effective star formation efficiency (column 2), the fraction of SNII energy that couples to gas (column 3), whether the model includes the impact of dust attenuation (column 4), the escape fraction of ionizing photons (column 5), the impact of X-ray heating on the IGM (column 6) and the escape fraction of Lyman Alpha photons (column 7). }
\centerlast
\begin{tabular}{|c| c| c| c| c| c |c|} 
 \hline
 model name & $f^{\rm eff}_{*}$ & $f_w$ & dust  & $f_{\rm esc}$  & $f_{\rm X,h}$ & $f_\alpha$ \\ [0.5ex] 
 \hline
 fiducial & 0.15 & 0.06 & yes & 0.1 & 0.2 & 1.0 \\ 
 intrinsic & 0.15 & 0.06 & no & 0.1 & 0.2 & 1.0\\
 maximal & 1.0 & 0.0 & no & 0.003 & 0.2 & 1.0\\
 \hline
\end{tabular}
\label{table1}
\end{table}

We compute the redshift evolution of the globally averaged mean 21cm differential brightness temperature ($ T_{b}$) at $z \sim 6-20$ following the calculations outlined in previous works \citep{furlanetto2006c, pritchard2012, 2020MNRAS.496.1445C}. For the cosmological parameters considered in this paper \citep[][]{planck2016}, at an observational frequency $\nu$, $T_{b}$ can be expressed as 
    \begin{equation}
         T_b(\nu) = 10.1~{\rm mK}~ x_{\rm HI}(z)~\left(1 - \frac{T_{\gamma}(z)}{T_S(z)}\right)~(1 + z)^{1/2},
        \label{eq:T_b}
    \end{equation}
    where $T_S$ is the spin temperature of neutral hydrogen, $T_{\gamma}$ is the Cosmic Microwave Background (CMB) temperature, and $x\rm_{HI}$ denotes the neutral hydrogen fraction present in the IGM at $z$. The spin temperature $T_S$ is calculated as \citep{1958PIRE...46..240F}
    \begin{equation}
    T_S^{-1}=\frac{T_{\gamma}^{-1}+x_{\alpha}T_{\alpha}^{-1}+ x_c T^{-1}_{K}}{1+ x_{c} + x_{\alpha}},
    \label{eq:tspin_without_assumption}
    \end{equation}
    
    where $x_{\alpha}$, $x_{c}$ are the Lyman Alpha (Ly$\alpha$) and the collisional coupling coefficients, respectively. The kinetic temperature of the IGM and  the color temperature associated with the Ly$\alpha$ background are denoted by $T_{K}$ and $T_{\alpha}$, respectively. In the redshift range of interest, i.e., $z=20-5$, the collisional coefficient is not expected to play a significant role \citep{pritchard2012}. Moreover, the high optical depth values for Ly$\alpha$ photons in this redshift range results in $T_{k} = T_{\alpha}$ \citep{pritchard2012}. Under these assumption the above equation simplifies to
    \citep{2021MNRAS.507.2405C}
    \begin{equation}                            T_S^{-1}=\frac{T_{\gamma}^{-1}+x_{\alpha}T_{K}^{-1}}{1+ x_{\alpha}},
    \label{eq:tspin}
    \end{equation}
    
    The redshift evolution of the IGM kinetic temperature ($T_{K}$) is primarily\footnote{Additional sources responsible for IGM heating include a (sufficiently) steep radio photon spectrum \citep{2023MNRAS.523.1908A}, cosmic ray photons \citep{2023arXiv230407201G, 2019MNRAS.483.5329J} and Ly$\alpha$ heating \citep{2021MNRAS.503.4264M}. However, in the interest of simplicity, we have ignored these terms in this work.} determined by the two key processes i.e., adiabatic cooling due to the expansion of the Universe and the X-ray heating of the IGM. The X-ray heating can be obtained from the X-ray emissivity as \citep{2012MNRAS.419.2095M} 
    \begin{equation}
        \frac{\epsilon_X(z)}{\rm J ~s^{-1} ~Mpc^{-3}} = 3.4 \times 10^{33} \left(
        f_{X} \times f_{h}\frac{{\rho}_{\rm SFRD}(z)}{\rm{\msun~yr^{-1}~Mpc^{-3}}}\right),
    \end{equation}
    where ${\rho}_{\rm SFRD}(z)$ is the redshift evolution of the star formation rate density as obtained from the \code{delphi} model. The term $f_{X}$ is an unknown efficiency parameter that is effectively a normalisation of the $\epsilon_{X}-\rho_{\rm SFRD}$ relation as compared to the local Universe, i.e. $f_{X}=1$ assumes high-$z$ X-ray sources to behave as in the local Universe. Further, $f_{h}$ denotes the fraction of the X-ray photons that go into heating the IGM, with the rest ionizing the IGM. We combine these two free parameters into one as $f_{X, h}= f_{X} \times f_{h}$ - this is the first free parameter for the 21cm calculations. For our fiducial model, we take $f_{X,h}=0.2$ \citep{furlanetto2006}, consistent with local Universe observations.
    
    The Ly$\alpha$ coupling coefficient ($x_{\alpha}$) is solely determined by the background Ly$\alpha$ flux, $J_{\alpha}$ \citep{2004ApJ...602....1C} which is computed as \citep{2003ApJ...596....1C}
    \begin{equation}
        J_{\alpha}=\frac{c}{4 \pi}(1+z)^3 \int^{z_{\rm max}}_{z}\dot{n}_{\nu'}(z') \left|\frac{d t'}{d z'}\right| dz'.
    \end{equation}
    Here $c$ is the speed of light and $t'$ is the cosmic time corresponding to the redshift $z'$. The upper limit to the integration, $z_{\rm max}$, has been calculated as $z_{\rm max} = (\nu_{LL}/\nu_{\alpha})(1+z)$ \citep{2020MNRAS.496.1445C} 
    where $\nu_{LL}$ denotes the Lyman limit frequency. Furthermore, $\nu' = \nu_{\alpha} (1 + z') / (1 + z)$, with $\nu_{\alpha}$ denoting the Ly$\alpha$ frequency. The term $\dot{n}_{\nu'}(z')$ is the rate of production of Ly$\alpha$ photons per unit frequency per unit comoving volume at redshift $z'$ emerging out of the galaxy, as obtained from the \code{delphi} model. This accounts for the fact that of the Ly$\alpha$ photons produced within a galaxy, only a fraction $f_\alpha$ emerge into the IGM - this is the second free parameter for the 21cm calculations. While we assume $f_\alpha=1$ for our fiducial model, we explore a range of parameters as described in what follows.

    Finally, the redshift evolution of the neutral hydrogen fraction, $x_{\rm HI}$ (used in Eqn. \ref{eq:T_b}) can be determined as
        \begin{equation}
        \frac{\der x_{\rm HI}}{\der t} = -f_{\rm esc}\frac{\dot{n}_{\rm ion}}{n_{H, {\rm com}}} + (1 - x_{\rm HI})~ \alpha_{B}~{\cal C}~n_{H,{\rm com}}~(1+z)^{3},
    \end{equation}
    where $\dot{n}_{\rm ion}$ is the production rate density of  ionizing photon in early galaxies as obtained from \code{delphi}, $n_{H, {\rm com}}$ is the hydrogen comoving number density, $f_{\rm esc}$ is the escape fraction of hydrogen ionizing photons, ${\cal C}$ is the clumping factor of the IGM and $\alpha_B$ is the (case B) recombination rate coefficient. The functional form of clumping factor ${\cal C}$ is taken as $1+43z^{-1.71}$ \citep{pawlik2009}. Further, we fix the value of $f_{\rm esc}$ for each model (intrinsic, fiducial and maximal) by matching to the CMB optical depth of $\tau = 0.055 \pm 0.007 $ \citep[][]{planck2016}. While we find $f_{\rm esc}=0.1$ \citep[consistent with a number of previous works, e.g.,] []{dayal2020,trebitsch2022,2023MNRAS.tmpL..54M} for both the fiducial and the intrinsic models, with its much higher star formation rate density (and hence the production rate of ionizing photons), the maximal model requires a much lower value of $f_{\rm esc}=0.003$ to match to the observed $\tau$ value.

\section{The redshift evolution of the global 21cm signal}
\begin{figure}
    \centering
    \includegraphics[width=0.5\textwidth]{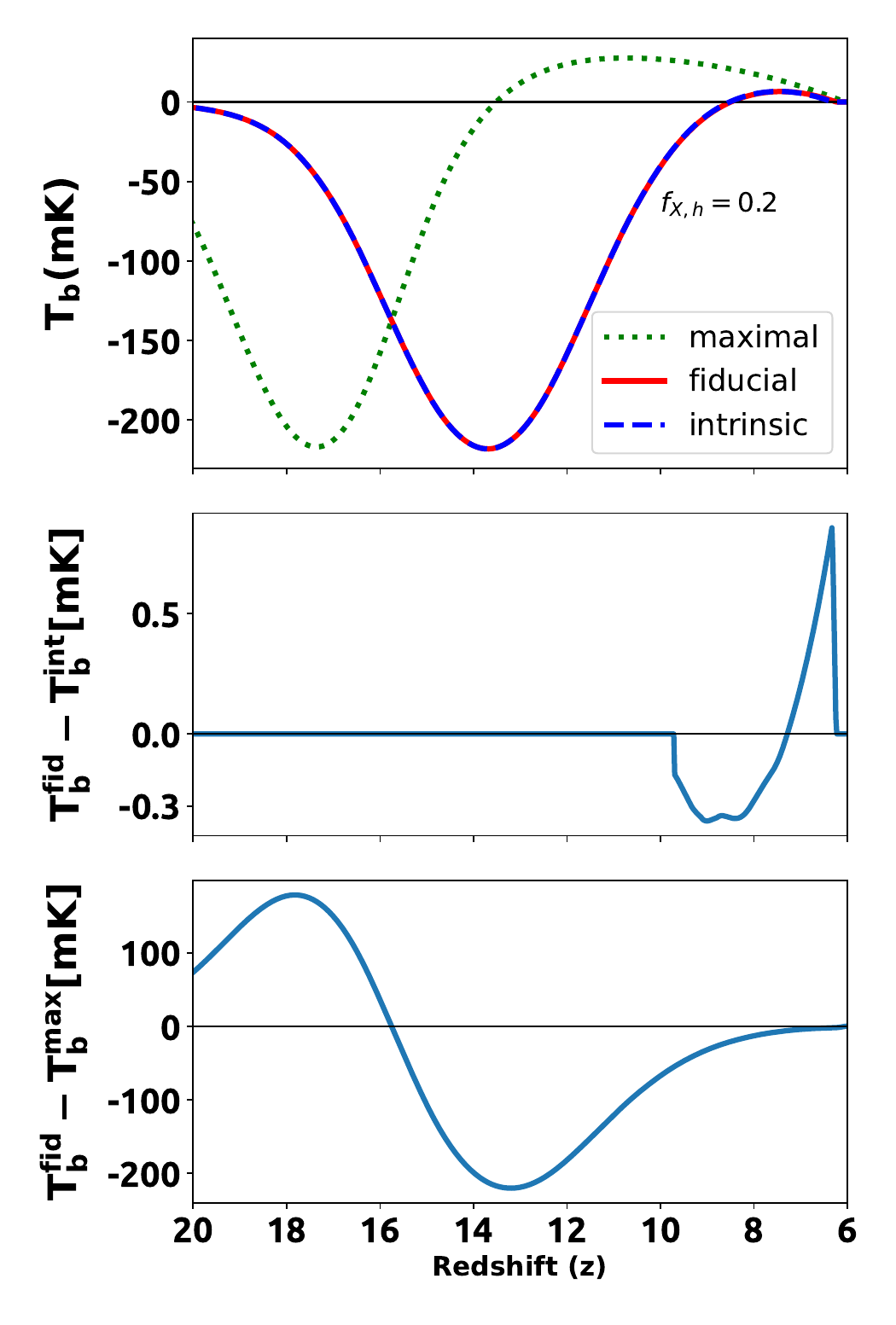}
    \caption{The {\it top panel} shows the redshift evolution of the global 21cm signal calculated using the intrinsic (dashed blue), fiducial (solid red) and maximal scenarios (dotted green) for the \code{delphi} semi-analytical model. For all these models, $f_{X,h}=0.2$ and $f_\alpha=1.0$. The {\it middle panel} shows the difference between the 21cm signal from the fiducial and intrinsic models, which, as expected, only differ slightly at $z \lsim 8$ where dust leads to a difference in the UV luminosity density obtained from these two models. Finally, the {\it bottom panel} shows the difference between the 21cm signal calculated from the fiducial and the maximal models.}
    \label{fig:delta_T_b}
\end{figure}

We now present the redshift evolution of the global 21cm signal for the fiducial, intrinsic and maximal galaxy formation models, as shown in Fig. \ref{fig:delta_T_b}. The 21cm signal computed from all of these models assumes $f_{X,h}=0.2$ and $f_{\alpha}=1.0$, as also specified in Table \ref{table1}.  

As seen from Fig. \ref{fig:uvld}, dust only has an impact on the UV luminosity density (and the associated Ly$\alpha$ luminosity and ionizing photon production rate density) at $z \lsim 8$. As a result, the intrinsic (no dust attenuation) and fiducial (with dust attenuation) models effectively predict the same 21cm signal at $z \sim 20-6$ as shown in the top panel of Fig. \ref{fig:delta_T_b}. For both models, we find a minimum brightness temperature of $ T_{\rm b, min}\sim -215$ mK at a redshift $z_{\rm min} \sim 14$. The global 21cm signal obtained from these two models, shown in the middle panel of the same figure, starts showing a small difference ($\sim 0.3-0.6$ mK) at $z \sim 10-6.5$ which can be explained as follows: the slightly larger star formation rate density in the intrinsic model (Fig. \ref{fig:uvld}) leads to a correspondingly higher value of the X-ray emissivity. The resulting higher spin temperature leads to a higher value of $T_{\rm b}$, i.e. we find $T^{\rm fid}_{\rm b} - T^{\rm int}_{\rm b} < 0$ at $z \sim 10-7$. At $z \lsim 7$, the evolution of the neutral hydrogen fraction starts dominating the brightness temperature equation. As $T_{S}$ increases with decreasing redshift, $T_{\gamma}/T_{S} << 1$ i.e. Eqn. \ref{eq:T_b} becomes independent of $T_{S}$. Given that we assume ionizing photons to be affected by dust in the same way as UV photons, the ``escaping" rate of ionizing photons in the intrinsic model is slightly more than that in the fiducial model at $z \lsim 8$. This results in a smaller value of $x_{\rm HI}$ in the intrinsic model, due to which $T^{\rm fid}_{b}$ starts to become larger than in the fiducial model, leading to $ T^{\rm fid}_{\rm b} - T^{\rm int}_{\rm b} > 0$. This difference disappears at $z \sim 6$ when reionization is completed, and the signals from both models converge to zero.

\begin{figure}
    \centering
    \includegraphics[width=0.5\textwidth]{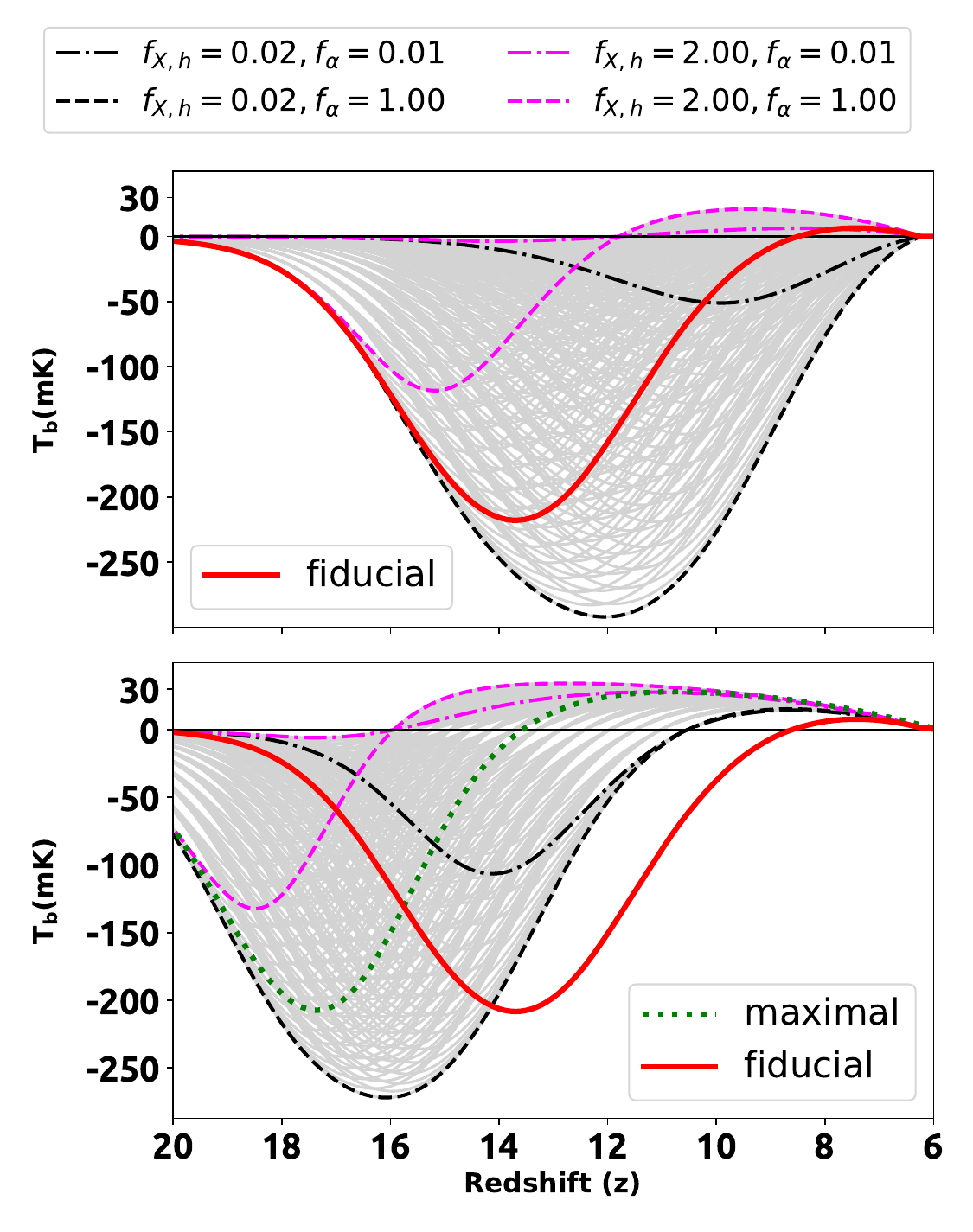}
    \caption{The {\it top and bottom panels} show the redshift evolution of the global 21cm signal for the fiducial (solid red) and maximal (dotted green) models in \code{delphi}, respectively. In each panel, the different grey curves show the signals obtained from 200 different combinations of $f_{X,h}$ and $f_{\alpha}$. We show 4 specific combinations, varying both $f_{X,h}$ and $f_\alpha$ by a factor of 10 around the fiducial model values, as marked, to show their impact on the global 21cm signal. }
    \label{fig:vary_parameter_max_fid}
\end{figure}

 We then focus on the 21cm signal obtained from the ``maximal" model. As discussed in Sec. \ref{sam}, this model presents an upper limit to the allowed UV luminosity density (and the associated Ly$\alpha$ production rate density and X-ray emissivity). As a result, in this model, the Ly$\alpha$ background saturates as early as $z\sim20$ showing $ T_{\rm b, min}= -210$ mK as early as $z_{\rm min} \sim 18$. The bottom panel of Fig. \ref{fig:delta_T_b} shows the difference between the 21cm signals produced from the fiducial and the maximal models. After the 21cm signal in the maximal model reaches its minima, X-ray heating starts, and the amplitude of the signal starts to increase, decreasing the difference between the maximal and fiducial model brightness temperatures. However, at $z \sim 13$, the signal from the fiducial model reaches its minimum and starts to increase. From this time onward, the difference between these two signals tends to go towards zero and eventually, at $z=6$, the difference disappears.

Considering the highly uncertain nature of the free parameters for the 21cm calculations ($f_{X,h}$ and $f_{\alpha}$), we then explore the physically plausible parameter space allowed  both in order to understand their impact on $z_{\rm min}$ and $ T_{\rm b, min}$ and to interpret the signal from forthcoming 21cm experiments. We compute the 21cm signal on a two-dimensional grid in $f_{X,h}=0.02-2.0$ and $f_{\alpha} = 0.01-0.1$ (exploring 200 combinations) for both the fiducial and maximal models, the results of which are shown in Fig. \ref{fig:vary_parameter_max_fid}. To highlight the impact of these free parameters on the 21cm signal, we focus on four combinations where $f_{X,h}=0.02 -2.0$ and $f_{\alpha}=0.01-1.0$, as shown in the same figure. 

Globally, the trends we find are the following: for a fixed value of $f_\alpha$\footnote{Once we fix $f_{\alpha}$, we are effectively fixing the redshift at which the spin temperature couples to the kinetic temperature.}, as $f_{X,h}$ increases from 0.02 to 2.0, the 21cm signal decreases in terms of its amplitude in addition to showing minima at earlier redshifts. As seen in the top panel, for the fiducial model, the minimum brightness temperature varies from as low as $T_{\rm b, min} \sim -290$ mK at $z_{\rm min} \sim 12$ (for $f_{X,h}=0.02$ and $f_{\alpha}=1.0$) to as high as $\sim -5$ mK ($f_{X,h}=2.0$ and $f_{\alpha}=0.01$) at $z_{\rm min} \sim 15$. This is driven by the fact that a higher value of $f_{X,h}$ causes the X-ray heating to start earlier, moving the absorption trough of the signal to appear at a higher redshift and causing a decrease in its amplitude. Further, for a fixed value of $f_{X,h}=0.02$, as $f_\alpha$ increases by a factor 100 from 0.01 to 1.0, the 21cm signal shows a much lower minimum value at increasingly higher redshifts - for example, for $f_{\alpha}=0.01$ $T_{b, min} =-51$ mK at $z_{\rm min} \sim 10$, whereas $T_{b, min} = -292$ mK for $f_{\alpha}=1.0$ at $z_{\rm min} \sim 12$. This is because the higher the value of $f_{\alpha}$, the earlier the redshift at which the Ly$\alpha$ background saturates and couples the spin temperature to the kinetic temperature. Given that the kinetic temperature will be lower at earlier redshifts, an increase in $f_{\alpha}$ leads the absorption trough to appear at an earlier redshift with a lower minimum value. In terms of comparison to observations, we note that while the minimum value of $T_{\rm b, min} \sim -290$ mK obtained from the fiducial model is consistent with the SARAS-3 non-detection (the reported RMS noise of the SARAS 3 measurements is 213 mK), obtaining a brightness temperature as low as that reported by EDGES ($-500\pm200$ mK) will require additional physics. 

As might be expected, we find the same qualitative trends from the maximal model (bottom panel of Fig. \ref{fig:vary_parameter_max_fid}), although as noted above, the 21cm signal here reaches its minimum at a much higher redshift of 18. In this case, we find $T_{\rm b, min}$ to vary between $ \sim -272$ mK to $-5.85$ mK, whereas $z_{\rm min}$ varies between $z_{\rm min} = 18 - 14$.

As seen from this figure, we also find a degeneracy between {\it galaxy formation} and {\it 21cm free parameters} with the fiducial and maximal models showing a minimum in the 21cm signal at very similar redshifts ($z\sim 15$) for different combinations of $f_{X,h}$ and $f_\alpha$. For example, the fiducial model with $f_{X,h}=0.2$ and $f_\alpha=1$ shows a very similar redshift behaviour compared to the maximal model with $f_{X,h}=0.02$ and $f_\alpha=0.01$. In this case, the higher star formation rate densities and X-ray emissivities in the maximal model are compensated by lower coupling parameters. However, the amplitude of the signal is much lower in the fiducial model ($T_{\rm b, min}= -210 $ mK) as compared to the maximal model ($T_{\rm b, min}= -106$ mK). Therefore, a combination of the minimum temperature and its redshift will be crucial in constraining galaxy formation parameters and their coupling to the 21cm signal at these early epochs.

\section{Conclusions and discussion} 
In this work, we calculate the global 21cm signal in the first billion years. The key strength of this work lies in the fact that the properties of our source galaxy population, obtained from the \code{delphi} model, are fully calibrated against the latest data sets from JWST and ALMA using only two redshift- and mass-independent free parameters. While our fiducial model well reproduces the galaxy population at $z \sim 5-12$ (using similar luminosity cuts as the observations), tentative photometric selections at $z \gsim 12$ seem to indicate extremely high values of the UV luminosity density for which we also calculate the 21cm signal using a ``maximal" model where each galaxy can convert 100\% of its gas into stars and there is no impact of SNII feedback. Our key findings are:
\begin{itemize}

\item Starting with a brightness temperature of about 0 mK at $z \sim 20$, the fiducial model (including dust attenuation) predicts a global 21cm signal whose amplitude decreases with decreasing redshift, reaching a minimum brightness temperature of $ T_{\rm b, min}\sim -215$ mK at a redshift $z_{\rm min} \sim 14$. The amplitude of the signal increases at $z \sim 13$ once X-ray heating starts becoming effective and reappears in emission between $z \sim 8-6$; the signal disappears at $z \sim 6$ when reionization completes.

\item We find that the inclusion of dust does not have a sensible impact on the global 21cm signal. Given that the impact of dust only becomes relevant at $z \lsim 8$, both the intrinsic (no dust attenuation) and fiducial (with dust attenuation) models effectively predict the same global 21cm signal at all $z \sim 20-6$.

\item The global 21cm signal from the ``maximal" model (with its star formation efficiency of 100\% and no SNII feedback) is qualitatively similar to that from the fiducial model. However, the higher star formation rates (and hence Ly$\alpha$ production) result in $ T_{\rm b, min}= -210$ mK as early as $z_{\rm min} \sim 18$.  

\item We also highlight a degeneracy between {\it galaxy formation} and {\it 21cm free parameters}. For example, the fiducial model with $f_{X,h}=0.2$ and $f_\alpha=1$ shows a very similar redshift behaviour compared to the maximal model with $f_{X,h}=0.02$ and $f_\alpha=0.01$; this is driven by the lower 21cm-coupling parameters compensating for the higher star formation rate densities and X-ray emissivities in the maximal model. A combination of the minimum temperature and its redshift will therefore be crucial in constraining galaxy formation parameters and their coupling to the 21cm signal at these early epochs.

\end{itemize}

Over the next years, the JWST will be crucial in confirming the nature of the (tentative) ultra high-redshift candidates detected out to $z \sim 18$ with ALMA providing further unrivalled constraints on the dust-obscured star formation rate density well within the first billion years. Further, even existing upper limits on the 21cm global signal in the redshift range $15-6$ will be crucial in obtaining constraints on the $f_{X,h}-f_{\alpha}$ parameter space: for example, the tentative current upper limits from the EDGES high band survey tend to disfavour 21cm signal with $ T_{\rm b, min}  < -200$ mK \citep{2018ApJ...863...11M} in the redshift range 15-6, which can effectively rule out some combinations of $f_{X,h}- f_{\alpha}$ with high $f_{\alpha}$ and low $f_{X,h}$, such as $f_{\alpha} = 1$ and $f_{X,h}= 0.02$.

\section*{Acknowledgments} 
AC wishes to acknowledge the computing facility provided by IUCAA. PD and VM acknowledge support from the NWO grant 016.VIDI.189.162 (``ODIN"). PD warmly thanks the European Commission's and University of Groningen's CO-FUND Rosalind Franklin program.

\section*{Data Availability}
Data generated in this research will be shared on reasonable request to the corresponding author.



\bibliographystyle{mnras}
\bibliography{main} 


\bsp	
\label{lastpage}
\end{document}